\documentclass[aps,prb,twocolumn,notitlepage,superscriptaddress]{revtex4-2}
\usepackage{amsmath}
\usepackage{amssymb}
\usepackage{graphicx}
\usepackage{epstopdf}
\usepackage[colorlinks=true]{hyperref}
\usepackage{mathtools}
\usepackage{cleveref}
\usepackage{soul}
\usepackage{booktabs}

\begin{document}


\title{The domain-wall motion driven by a rotating field in a ferrimagnet}


\author{Munsu Jin}
\thanks{These authors contributed equally to this work.}
\affiliation{Department of Physics, Korea Advanced Institute of Science and Technology, Daejeon 34141, Republic of Korea}
\author{Ik-Sun Hong}
\thanks{These authors contributed equally to this work.}
\affiliation{KU-KIST Graduate School of Converging Science and Technology, Korea University, Seoul 02841, Republic of Korea}
\author{Duck-Ho Kim}
\affiliation{Center for Spintronics, Korea Institute of Science and Technology, Seoul 136-791, Republic of Korea}
\author{Kyung-Jin Lee}
\affiliation{Department of Physics, Korea Advanced Institute of Science and Technology, Daejeon 34141, Republic of Korea}

 \author{Se Kwon Kim}
 \email{sekwonkim@kaist.ac.kr}
\affiliation{Department of Physics, Korea Advanced Institute of Science and Technology, Daejeon 34141, Republic of Korea}


\date{\today}

\begin{abstract}
We theoretically study a ferrimagnetic domain-wall motion driven by a rotating magnetic field. We find that, depending on the magnitude and the frequency of the rotating field, the dynamics of a ferrimagnetic domain wall can be classified into two regimes. First, when the frequency is lower than a certain critical frequency set by the field magnitude, there is a stationary solution for the domain-wall dynamics, where a domain-wall in-plane magnetization rotates in-phase with the external field. The field-induced precession of the domain wall gives rise to the translational motion of the domain wall via the gyrotropic coupling between the domain-wall angle and position. In this so-called phase-locking regime, a domain-wall velocity increases as the frequency increases. Second, when the frequency exceeds the critical frequency, a domain-wall angle precession is not synchronous with the applied field. In this phase-unlocking regime, a domain wall velocity decreases as the frequency increases. Moreover, the direction of the domain-wall motion is found to be reversed across the angular compensation point where the net spin density of the ferrimagnet changes its sign. Our work suggests that the dynamics of magnetic solitons under time-varying biases may serve as platform to study critical phenomena.
\end{abstract}


\maketitle

\section{Introduction}

Spintronics is the field which aims at advancing information technology beyond what has been achievable with charge-based electronics by exploiting spin degree of freedom~\cite{ZuticRMP2004}. A natural venue to look for spin-based functionality is magnet materials, which are known to exhibit various excitations that can be used for information carriers such as spin waves and topological solitons~\cite{ChumakNP2015, kosevich1990magnetic}. In particular, a magnetic domain wall, which is a prototypical soliton in easy-axis magnets, has been a subject of intensive studies in spintronics due to its technological utilities as topologically robust information carriers as well as intriguing physics~\cite{allwood2005magnetic, parkin2008magnetic}. For example, a domain-wall racetrack memory, where a series of domain walls are moved along the one-dimensional racetrack while carrying information, has been shown to have potential for fast, nonvolatile, and three-dimensional solid-state memory architecture~\cite{parkin2008magnetic, ParkinNN2015}. In addition to practical utilities, a domain wall is known to exhibit various fundamentally interesting nonlinear phenomena. One example is given by the so-called Walker breakdown, which refers to a phenomenon of sudden drop of the domain-wall velocity at certain critical strength of the driving force due to the onset of the precession motion of domain wall~\cite{schryer1974motion}. The Walker breakdown in the field-driven domain-wall motion has been experimentally demonstrated in ferromagnetic wires, e.g., in Ref.~\cite{beach2005dynamics}. 

In efforts to expand material platforms for spintronics from ferromagnets that have been conventional material platform for spintronics, antiferromagnets have been receiving much attention in spintronics as alternative material choices due to their certain advantages over ferromagnets~\cite{kimel2009inertia, loth2012bistability, jungwirth2016antiferromagnetic}. For example, the dynamics of antiferromagnets are known to exhibit THz intrinsic frequency, which is generally faster than ferromagnetic dynamics which is on the order of GHz. Also, the absence of the equilibrium magnetization of antiferromagnets allows for the development of denser spintronic devices compared to ferromagnet-based devices which suffer from strong cross-device interactions mediated by the stray field. In particular, antiferromangetic domain wall has been shown to be fundamentally different from ferromagnetic counterparts and thus has been studied intensively in the last decade. For example, antiferromagnetic domain wall is shown to not exhibit the Walker breakdown unlike a ferromagnetic case and thus can be driven with higher velocities~\cite{gomonay2016high}. Also, when the antiferromagnetic domain-wall velocity is close to the maximum magnon group velocity, it has been experimentally demonstrated to exhibit the Lorentz-like contraction by shrinking its width according to the relativistic kinematics~\cite{BaryakhtarJETP1983, HaldanePRL1983, BaryakhtarSPU1985, kim2014propulsion, shiino2016antiferromagnetic, CarettaScience2020}. Despite the fundamental interest and technological potentials, however, it is still experimentally challenging to detect and control antiferromagnetic dynamics due to its zero net magnetization, although there have been some progress enabled by x-ray absorption spectroscopy~\cite{weber2003magnetostrictive, salazar2009direct, wu2011direct}, spin-polarized scanning tunneling microscopy~\cite{ bode2006atomic, loth2012bistability}, and quantum sensing with single spins~\cite{ gross2017real, kosub2017purely}. 
 
Recently, ferrimagnets, which consist of two or more inequivalent magnetic sublattices that are coupled antiferromagnetically, have emerged in spintronics as material platforms that can offer advantages of both ferromagnets and antiferromagnets~\cite{FinleyAPL2020}. They generally have a small, but finite magnetization and thus can be detected and controlled by conventional methods used for ferromagnets. Also, under suitable conditions, their dynamics resembles the dynamics of antiferromagnets since their magnetic sublattices are antiferromagnetically coupled similarly to antiferromagnets~\cite{kirilyuk2010ultrafast}. In other words, antiferromagnet-like dynamics of ferrimagnets is controllable and detectable due to its finite magnetization. This feature of ferrimangets, typified by rare-earth transition-metal (RE-TM) ferrimagnets, allows for the fast domain-wall motion~\cite{kim2017, caretta2018fast, siddiqui2018current, cai2020ultrafast} and ultrafast magnetization switching~\cite{ostler2012ultrafast, finley2016spin, mishra2017anomalous}. In this work, we are interested in the dynamics of a domain wall motion in a ferrimagnet.

Most of the studies on a domain-wall motion have focused on the effects of DC biases such as an external field and a current. In searching for novel magnetic phenomena, domain-wall motion by oscillating biases have been receiving increasing attention in the field. For example, the motion of an antiferromagnetic domain wall by a rotating field has been studied in Refs.~\cite{pan2018driving, li2020rotating}. Also, it has been shown that a ferromagnetic domain-wall motion driven by time-periodic field or current can exhibit a sudden drop of the domain-wall velocity akin to the Walker breakdown~\cite{tatara2004theory, kim2020magnetic}. In Ref.~\cite{kim2020magnetic}, this Walker-like breakdown of an AC-bias-driven ferromagnetic domain-wall motion has been explained by phase-locking and phase-unlocking transition, which mimics an analogous phenomenon in an electric RLC circuit discovered by Adler~\cite{adler1946study}. In spintronics, the phase-locking of the spin-torque oscillator to an AC current has also been explained by invoking its analogy to the Adler equation~\cite{rippard2005injection, georges2008coupling, urazhdin2010fractional}. Although the AC-bias-driven domain-wall motion has been studied for ferromagnets and antiferromagnets, the corresponding problem for a ferrimagnetic domain wall has not been studied.

\begin{figure}
\includegraphics[width=\columnwidth]{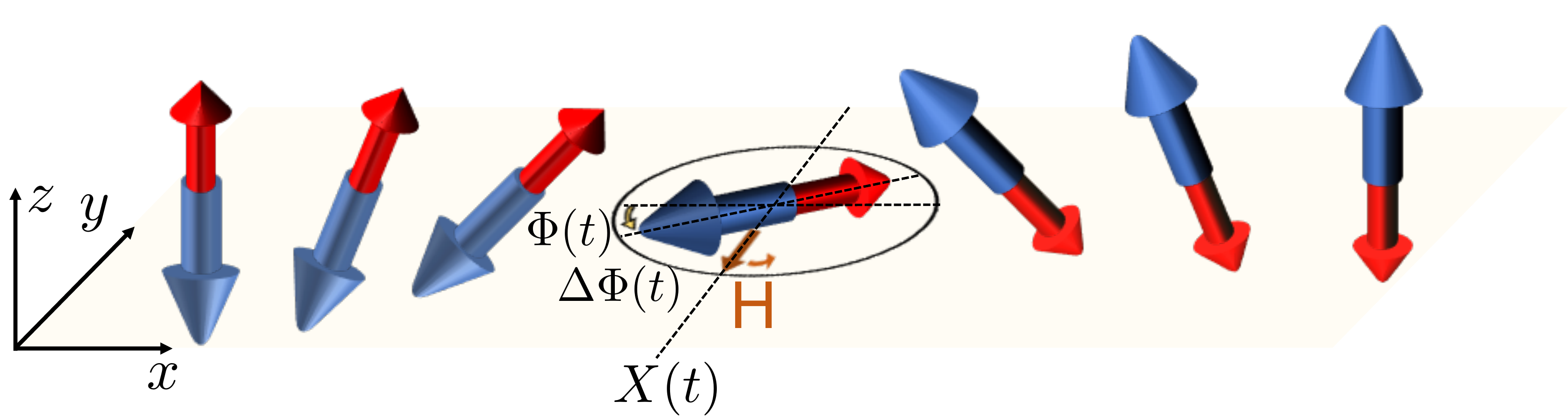}
\caption{Schematic of the magnetization configuration of a ferrimagnetic domain wall, where blue and red arrows represent the magnetic moments of two sublattices of the ferrimagnet. The domain-wall position is denoted by $X(t)$ and the in-plane domain-wall magnetization angle is denoted by $\Phi(t)$. The external field $\mathbf{H}(t)$ rotates within the $xy$ plane and the domain-wall in-plane magnetization lags behind the external field by $\Delta \Phi(t)$.}
\label{fig:fig1}
\end{figure}

In this paper, we theoretically investigate a ferrimagnetic domain-wall motion driven by a rotating field, which is schematically illustrated in Fig.~\ref{fig:fig1}. We find that, when the frequency is below a certain critical frequency, the precession of the in-plane magnetization inside the domain wall is synchronized with the applied rotating field. In this low-frequency regime, the domain-wall velocity increases linearly as the frequency increases. When the frequency exceeds the critical frequency, on the other hand, the domain-wall motion cannot keep pace with the rotating field, making its motion asynchronous with the rotating field. In this case, the domain-wall velocity decreases as the frequency increases. We refer the former and the latter regimes as the phase-locking regime and the phase-unlocking regime, respectively. The analytical solutions are checked by performing the numerical simulations, which show good agreement in both phase-locking and phase-unlocking regimes. The unique feature of the ferrimagnetic domain-wall motion occurs in the vicinity of angular momentum compensation point ($T_A$): the direction of the domain-wall motion is reversed as the ferrimagnet passes across $T_A$ due to the sign flip of the net spin density at $T_A$. For experimental feasibility of using a rotating field to drive a domain wall, we would like to mention that there has already been an experimental demonstration of the chirality reversal of a vortex domain wall induced by a rotating magnetic field~\cite{bisig2015dynamic}.

This paper is organized as follows. In Sec.~\ref{sec2}, we develop a theory for the dynamics of a ferrimagnetic domain wall in the presence of a rotating field within the Landau-Lifshitz-Gilbert-like equations of motion for ferrimagnets. The main analytical results are the critical frequency [Eq.~(\ref{eq:wh})] that separates between the phase-locking and the phase-unlocking regimes, the domain-wall velocity as a function of the frequency in the phase-locking regime [Eq.~(\ref{eq:V1})] and the velocity in the phase-unlocking regime [Eq.~(\ref{eq:V2})]. In Sec.~\ref{sec3}, we present our numerical simulation results and compare them with the analytical solutions. In Sec.~\ref{sec4}, we summarize our work.

\section{Theory for domain-wall dynamics driven by a rotating field}
\label{sec2}

In this section, we develop a theory for the dynamics of a ferrimagnetic domain wall driven by a rotating field in close connection to the existing theory for the dynamics of a ferromagnetic domain wall driven by an AC field~\cite{kim2020magnetic}. For concreteness, we consider a RE-TM ferrimagnets, where RE magnetic moments and TM magnetic moments are exchange-coupled antiferromagnetically.

\subsection{Analytical model}

The dynamics of ferrimagnets is described by the Landau-Lifshitz-Gilbert (LLG)-like equation, which is given by~\cite{ivanov1983nonlinear, bar1985dynamics,chiolero1997macroscopic, kim2017self, tserkovnyak2008theory, okuno2020spin,ivanov2019ultrafast}
\begin{equation}
\delta_s\dot{\textbf{n}}- \alpha s \textbf{n}\times\dot{\textbf{n}}-\rho\textbf{n}\times\ddot{\textbf{n}}=-\textbf{n}\times\textbf{h}_{\text{eff}} \, ,
\end{equation}
where $\textbf{n}$ is the unit vector in the direction of the magnetization of the TM sublattice, $\delta_s = s_\text{TM} - s_\text{RE}$ is the net spin density of a ferrimagnet along $\textbf{n}$, $s = s_\text{TM} + s_\text{RE}$ is the sum of spin densities of two sublattices, $\alpha > 0$ is the Gilbert damping constant, $\textbf{h}_{\text{eff}}\equiv -\delta U/\delta \textbf{n}$ is the field conjugate to the order parameter $\mathbf{n}$, and $\rho$ is the moment of inertia of the antiferromagnetic dynamics for the staggered magnetization $\textbf{n}$~\cite{jungwirth2016antiferromagnetic}, which is inversely proportional to the microscopic exchange energy between the two magnetic sublattices. 

We consider a quasi-one-dimensional ferrimagnet in a rotating field, which can be modeled by the potential energy $U=\int dx [\{ A (\partial_x\textbf{n})^2 - K (n_z)^2  + K_y(n_y)^2 \} / 2 -M \textbf{H}\cdotp\textbf{n}]$, where $A$ is the exchange coefficient, $K>0$ is the easy-axis anisotropy (also called perpendicular magnetic anisotropy), $K_y > 0$ is hard-axis anisotropy that captures the shape anisotropy induced by the magnetostatic interaction, $M = M_\text{TM} - M_\text{RE}$ is the net magnetization of the ferrimagnet, $\mathbf{H} = H (\cos(\omega t),\,\sin(\omega t),\,0)$ represents the rotating field about the $z$ axis at the frequency given by $\omega$. Without loss of generality, we consider the cases with $H > 0$ and $\omega > 0$. In this work, we neglect the nonlocal dipolar interaction since, due to the antiferromagnetic alignment of the two magnetic sublattices, the net magnetization of ferrimagnets is orders of magnitude smaller than that of ferromagnets.

Due to the easy-axis anisotropy, the ferrimagnet supports a stable nonlinear soliton solution with boundary condition $\textbf{n}(x\rightarrow\pm\infty)=\pm\hat{\textbf{z}}$, which is a called a domain wall. An equilibrium domain-wall solution is given by the following Walker ansatz~\cite{schryer1974motion}:
\begin{equation}
\mathllap{\textbf{n}}=\left(\cos\Phi \, \text{sech}\frac{x-X}{\lambda}, \sin \Phi \, \text{sech}\frac{x-X}{\lambda}, \text{tanh} \frac{x-X}{\lambda} \right) \, ,
\end{equation}
where $\lambda = \sqrt{A/K}$ is the parameter for the domain-wall width, $X$ represents the domain-wall position, and $\Phi$ is the in-plane angle of the domain-wall magnetization. See Fig.~\ref{fig:fig1} for the schematic illustration of the domain wall. The domain-wall position $X$ represents a zero-energy mode associated with the spontaneous breaking of the translational symmetry of the system. By plugging the domain-wall solution to the potential energy $U$, we obtain the following energy of the domain wall:
\begin{equation}
U(\Phi) = - \pi \lambda M H \cos (\Phi - \omega t) + \lambda K_y \sin^2 \Phi \, .
\label{eq:U}
\end{equation}
This result indicates that when the external field is sufficiently strong, $H \gg K_y / M$, the domain-wall angle $\Phi$ will follow $\omega t$, i.e., the phase of the external field, closely to minimize the Zeeman energy. When the anisotropy dominates the external field $K_y \gg M H$, the domain-wall angle will be kept closely to $0$ or $\pi$ and there would be no appreciable effect of the external field on the domain-wall dynamics.

The low-energy dynamics of the domain wall can be described the dynamics of the two collective coordinates, $X(t)$ and $\Phi(t)$. Within the collective-coordinate approach, we can derive the following coupled equations from the LLG-like equation~\cite{thiele1973steady, thiaville2005micromagnetic, tretiakov2008, kim2020dynamics}
\begin{equation}\label{eqn3}
-2\alpha s \dot{X}+2\delta_s\lambda\dot{\Phi}-2\rho\ddot{X}=0 \, ,
\end{equation}
and
\begin{align}\label{eqn4}
	\begin{aligned}
\mathllap{-}&2\alpha s \lambda \dot{\Phi}-2\delta_s\dot{X}-2\rho\lambda\ddot{\Phi}\\
&=2\lambda K_y\sin{\Phi}\cos{\Phi}+\pi\lambda M_s H\sin{(\Phi-\omega t)}.
	\end{aligned}
\end{align}
In this work, we are interested in the time-averaged dynamics of the domain wall over sufficiently long time. By taking time-average of Eq. (\ref{eqn3}), we obtain the following average domain-wall velocity:
\begin{equation}\label{gyro}
 \langle\dot{X}\rangle=\frac{\lambda \delta_s}{\alpha s}\langle\dot{\Phi}\rangle,
\end{equation}
where $\langle\ddot{X}\rangle$ and $\langle\ddot{\Phi}\rangle$ are set to be zero by assuming that the domain-wall dynamics is periodic such that the velocity $\dot{X}(t)$ and the angular velocity $\dot{\Phi}(t)$ are periodic functions of time $t$. The domain-wall velocity $\langle \dot{X} \rangle$ is linearly proportional to the angular precession of the magnetization $\langle \dot{\Phi} \rangle$, which is rooted in the gyrotropic coupling between $X$ and $\Phi$~\cite{tretiakov2008}. Note that the net spin density $\delta_s$ appears in the proportionality constant. In ferrimagnets, the value $\delta_s$ varies when the temperature changes. In particular, it changes sign across the angular momentum compensation point $T_A$, which will be invoked below to argue that, for the given rotating field, the sign of the domain-wall velocity flips as the temperature varies across $T_A$.

The coupling [Eq.~(\ref{gyro})] between $\langle \dot{X} \rangle$ and $\langle \dot{\Phi} \rangle$ enables us to drive the domain wall by a rotating field. For example, when the field magnitude $H$ is sufficiently strong and the field rotation is sufficiently slow, the in-plane magnetization inside the domain wall will be mostly parallel to the field direction $\mathbf{H}(t) = H (\cos (\omega t), \sin (\omega t), 0)$. This means that the domain-wall angle $\Phi$ follows $\omega t$ closely, leading to $\langle \dot{\Phi} \rangle \approx \omega$. Then, when the net spin density is finite $\delta_s \neq 0$, the domain wall should move at average velocity given by $\langle \dot{X} \rangle \approx \lambda \delta_s \omega / (\alpha s)$. Understanding the domain-wall dynamics for general situations, e.g., with higher frequencies, requires more sophisticated analysis, which we present below.

The time-averaged Eq.~(\ref{eqn4}) can be solely written in terms of the domain-wall angle $\Phi$ by replacing $\langle \dot{X} \rangle$ by $\frac{\lambda \delta_s}{\alpha s}\langle\dot{\Phi}\rangle$, which results in
\begin{equation}\label{49}
\langle\dot{\Phi}\rangle= - \omega_H\langle\sin{(\Phi-\omega t)}\rangle - \omega_K\langle\sin{2\Phi}\rangle \, ,
\end{equation}
where 
\begin{equation}
\label{eq:wh}
\omega_H\equiv\frac{\alpha s\pi M_s H}{2\{(\alpha s)^2+\delta_s^2\}} \, , 
\end{equation}
is the characteristic frequency determined by the external field and
\begin{equation}
\omega_K\equiv\frac{\alpha s K_y}{2\{(\alpha s)^2+\delta_s^2\}} \, ,
\end{equation}
is the characteristic frequency determined by the hard-axis anisotropy. This equation describes the dynamics of the domain-wall angle $\Phi$ driven by a rotating field in the $xy$ plane. In this work, we are interested in the effect of the rotating field on the domain-wall dynamics. Therefore, we will restrict our attention to the situations where the external field dominates the hard-axis anisotropy term so that we can set $K_y = 0$ and $\omega_K = 0$. With this approximation, Eq.~(\ref{49}) is reduced to $\langle \dot{\Phi} \rangle = - \omega_H \langle \sin{(\Phi-\omega t)} \rangle$. Instead of solving this averaged version, we will present an exact solution of the following equation
\begin{equation}\label{11}
\dot{\Phi} = \omega_H \sin{(\omega t-\Phi)} \, ,
\end{equation}
and will use the solution to obtain the domain-wall velocity $\langle \dot{X} \rangle$ as a function of the field magnitude $H$ and the field frequency $\omega$. Our analysis results, which will be obtained below, will be compared with the simulation results in Sec.~\ref{sec3}.

\subsection{Phase-locking and phase-unlocking regimes}

To solve Eq.~(\ref{11}), let us introduce a new parameter $\Delta\Phi\equiv\omega t-\Phi$. The physical meaning of $\Delta\Phi$ is the phase difference between the domain-wall angle and the rotating field. The equation of motion for $\Delta\Phi$ is given by
\begin{equation}\label{locking}
\frac{d\Delta\Phi}{dt}=\omega-\omega_H\sin{\Delta\Phi}.
\end{equation}
This equation has been studied in the field of nonlinear dynamics~\cite{Strogatz}. Note that two frequencies appear in the equation: the field rotation frequency $\omega$ and the magnitude-related frequency $\omega_H$. Depending on the relative magnitude of these two frequencies, the dynamics is divided into two regimes.

First, let us consider the cases where $\omega_H > \omega$, i.e., the cases where the field is sufficiently strong or the frequency is sufficiently slow.  In this regime, the equation permits a steady-state solution given by
\begin{equation}
\sin^{-1}\frac{\omega}{\omega_H}=\Delta\Phi \, , \quad \text{for } \omega < \omega_H \, .
\label{eq:sol1}
\end{equation}
In this regime, $\Delta \Phi$ is constant, which means that the domain-wall angle evolves in time while keeping its phase difference with the rotating field constant. Since the phase difference between the external field and the domain-wall angle is locked due to the strong external field, this regime is called a phase-locking regime. In this regime, according to Eq.~(\ref{gyro}), the average domain-wall velocity is given by
\begin{equation}
 \langle\dot{X}\rangle=\frac{\lambda \delta_s}{\alpha s}\omega \, , \quad \text{for } \omega < \omega_H \, .
 \label{eq:V1}
 \end{equation}
 This our first main result: The domain-wall velocity is linearly proportional to the rotating-field frequency for $\omega < \omega_H$. Note that the sign of the velocity depends on the sign of the net spin density $\delta_s$. Therefore, for the given rotating field, the domain-wall velocity should change its sign when $\delta_s$ changes the sign, i.e., when the temperature crosses the angular momentum compensation point $T_A$.
 
Second, let us consider the cases with $\omega > \omega_H$, where the rotation frequency is large compared to $\omega_H$. In this case, Eq.~(\ref{locking}) does not possess a steady-state solution. It still permits an exact solution given implicitly by 
 \begin{equation}
\tan\frac{\Delta\Phi}{2}=\frac{\omega_H}{\omega} + \sqrt{1-\frac{\omega_H^2}{\omega^2}}\tan{\frac{\sqrt{\omega^2-\omega_H^2}(t-t_0)}{2}} \, ,
\label{eq:sol2}
\end{equation} 
where $t_0$ is an arbitrary constant. Note that the period of the solution is given by $T = 2 \pi / \sqrt{\omega^2 - \omega_H^2}$, which is longer than the period of the applied field $2 \pi / \omega$, implying that the evolution of the domain-wall angle $\Phi$ is not in-phase with the rotating field. For this reason, the domain-wall dynamics with $\omega > \omega_H$ is referred to be in the phase-unlocking regime. The averaged angular velocity is given by $\langle \Delta\dot{\Phi} \rangle = 2 \pi / T = \sqrt{\omega^2 - \omega_H^2}$, and thus, from Eq.~(\ref{gyro}), the averaged domain-wall velocity is given by
\begin{equation}
\langle\dot{X}\rangle=\frac{\lambda \delta_s}{\alpha s} \left( \omega-\sqrt{\omega^2-\omega_H^2} \right) \, , \quad \text{for } \omega > \omega_H \, ,
\label{eq:V2}
\end{equation}
which is a decreasing function of $\omega$. This is our second main result: In the phase-unlocking regime ($\omega > \omega_H$) where the external field rotates too fast for the domain wall to keep pace with it, the average domain-wall velocity decreases as the frequency increases.

\begin{table*}[tbp]
\caption{Material parameters used for simulations. $M_\text{TM}$, $M_\text{RE}$, $s_\text{TM}$, $s_\text{RE}$, $\delta_s$, and $s$ are the magnetization of TM elements, the magnetization of RE elements, the spin density of TM elements, the spin density of RE elements, the net spin density, and the total spin density, respectively. $T_4$ and $T_7$ represent $T_M$ and $T_A$, respectively. \label{table}}
\begin{center}
	\resizebox{\textwidth}{!}{%
	\begin{tabular}{cccccccccc}
	\midrule[2pt]
Index & $T_1$ & $T_2$ & $T_3$ & $T_4$ ($T_M$)  & $T_5$ & $T_6$ & $T_7$ ($T_A$) & $T_8$ & $T_9$ \\\midrule[2pt]
$M_{TM}$ (emu/cm$^3$) & 1170 & 1140 & 1110 & 1080 & 1050 & 1020 & 990& 960 & 930\\\hline
$M_{RE}$ (emu/cm$^3$)& 1260& 1200& 1140& 1080 & 1020 & 960 & 900 & 840 & 780  \\\hline
$s_{TM}$ (erg$\cdot$s/cm$^3$)&$6.04\times10^{-5}$&$5.89\times10^{-5}$&$5.73\times10^{-5}$&$5.58\times10^{-5}$&$5.42\times10^{-5}$&$5.27\times10^{-5}$&$5.11\times10^{-5}$&$4.96\times10^{-5}$&$4.8\times10^{-5}$\\\hline
$s_{RE}$ (erg$\cdot$s/cm$^3$)&$7.16\times10^{-5}$&$6.82\times10^{-5}$&$6.48\times10^{-5}$&$6.14\times10^{-5}$ &$5.8\times10^{-5}$&$5.45\times10^{-5}$&$5.11\times10^{-5}$&$4.77\times10^{-5}$&$4.43\times10^{-5}$ \\\hline
$\delta_s(=s_A-s_B)$&$-1.1\times10^{-5}$&$-9.3\times10^{-6}$&$-7.4\times10^{-6}$&$-5.6\times10^{-6}$ &$-3.7\times10^{-6}$&$-1.9\times10^{-6}$&$0$&$1.86\times10^{-6}$&$3.72\times10^{-6}$ \\\hline
$s(=s_A+s_B)$&$1.3202\times10^{-4}$&$1.2707\times10^{-4}$&$1.2211\times10^{-4}$&$1.1715\times10^{-4}$&$1.1219\times10^{-4}$&$1.0723\times10^{-4}$&$1.0557\times10^{-4}$&$9.7314\times10^{-5}$&$9.2355\times10^{-5}$\\\hline
	\end{tabular}}
	\end{center}
\end{table*}

\section{Numerical analysis}
\label{sec3}

To confirm the analytical results, particularly the domain-wall velocity in the phase-locking regime [Eq.~(\ref{eq:V1})], and one in the phase-unlocking regime [Eq.~(\ref{eq:V2})], we performed the atomistic spin simulations by solving the two coupled Landau-Lifshitz-Gilbert equations for two antiferromagntically-coupled sublattices representing TM and RE magnetizations. The material parameters that used in the simulations are $A = 2.5 \times 10^{-7}$ erg/cm, $K = 9.5 \times 10^7$ erg/cm$^3$, and $K_y = 3 \times 10^3$ erg/cm$^3$, and $\alpha = 0.002$. The used gyromagnetic ratios of the TM and the RE sublattices are $\gamma_{TM}=1.936\times10^7$ s$^{-1}$Oe$^{-1}$ and $\gamma_{RE}=1.76\times10^7$ s$^{-1}$Oe$^{-1}$, respectively. The cell size was $0.4\times50\times1$ nm$^{3}$, corresponding to $x$, $y$, and $z$ axis, respectively. The system size is $400\times50\times1$ nm$^3$. Table~\ref{table} shows the magnetization and the spin density parameters that we used to model the effect of the temperature. The temperature $T_4$ corresponds to the magnetization compensation point $T_M$ where the magnetizations of the two sublattice are equal and thus the net magnetization vanishes. The temperature $T_7$ corresponds to the angular momentum compensation point $T_A$ where the spin densities of the two sublattices are equal and thus the net spin density vanishes. The applied field strengths are 3000 Oe for $T_1, T_2, T_3$, 1000 Oe for $T_4, T_5$, and 200 Oe for $T_6, T_7, T_8, T_9$. The field magnitude is chosen for each temperature such that the resultant domain-wall velocities are comparable.

 \begin{figure}
\includegraphics[width=0.7\columnwidth]{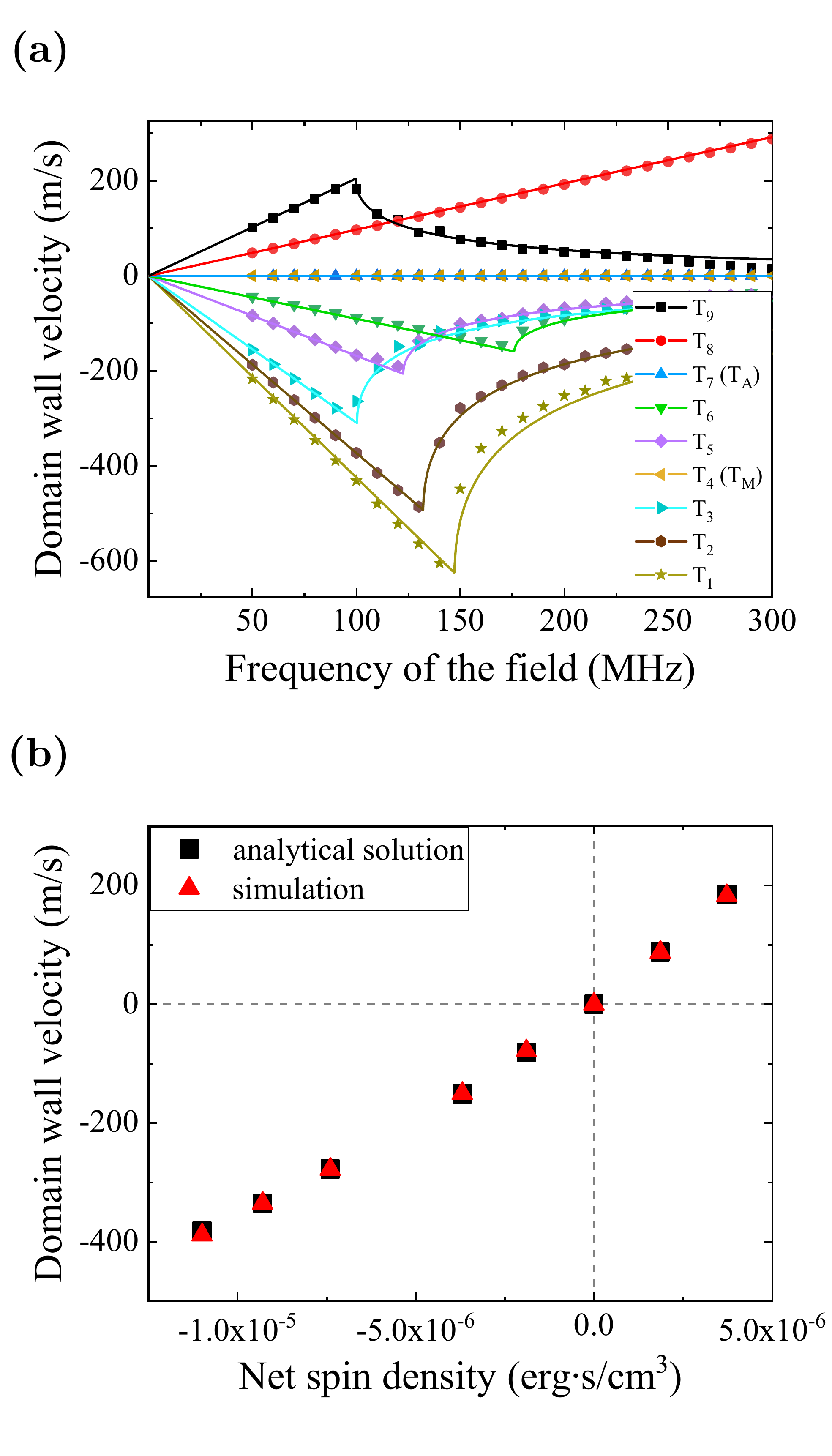}
\caption{(a) Domain-wall velocity $\langle \dot{X} \rangle$ as a function of the rotating-field frequency $f = \omega / 2 \pi$ for various configurations with the parameters shown in Table~\ref{table}. The lines are analytical solutions, Eq.~(\ref{eq:V1}) for $\omega < \omega_H$ (phase-locking regime where a domain-wall magnetization precesses at the same frequency as the external field) and Eq.~(\ref{eq:V2}) for $\omega > \omega_H$ (phase-unlocking regime where a domain-wall magnetization rotates slower than the external field). The dots represent simulation results. Note that the domain-wall velocity changes its sign as the temperature varies across the angular momentum compensation point $T_A$. (b) Domain-wall velocity as a function of the net spin density $\delta_s$ within the phase-locked regime for the frequency $f = 90$ MHz. The data from all the temperatures $T_1, T_2, \cdots, T_9$ except the magnetization compensation point $T_4 (T_M)$, where the frequency $f = 90$ MHz belongs to the phase-unlocking regime, is used. The squares and the triangles represent the analytical solutions [Eq.~(\ref{eq:V1})] and the simulation results, respectively.}
\label{fig:fig2}
\end{figure} 

Figure~\ref{fig:fig2}(a) shows the domain-wall velocity $\langle \dot{X} \rangle$ as a function of the frequency of the rotating field for various configurations. The lines represent the analytical results, which are given by Eq.~(\ref{eq:V1}) for $\omega < \omega_H$ and Eq.~(\ref{eq:V2}) for $\omega > \omega_H$ with the critical frequency $\omega_H$ given by Eq.~(\ref{eq:wh}). The dots represent the simulation results. The analytical results and the simulation results agree with each other reasonably well. Several features are noteworthy. First, at $T_M$ where the magnetization is zero, the domain-wall velocity vanishes, which is due to the absence of the coupling of the external field and the domain wall. Second, at $T_A$ where the net spin density is zero, the velocity vanishes. In our analytical results, the domain-wall velocity is proportional to $\delta_s$, and thus it is expected to vanish at $T_A$. Physically, this is due to the absence of the gyrotropic coupling between the domain-wall position $X$ and the domain-wall angle $\Phi$ at $T_A$. Thirdly, the sign of the domain-wall velocity, i.e., the direction of the domain-wall motion depends on the sign of the net spin density $\delta_s$. For $T_1, T_2, \cdots, T_6$ where $\delta_s < 0$, the sign of the velocity is negative, and for $T_8$ and $T_9$ where $\delta_s > 0$, the sign of the velocity is positive. This means that for the given rotating field, if we vary the temperature of the ferrimagnet across $T_A$, the direction of the domain-wall motion should reverse exactly at $T_A$, which may be exploited to detect $T_A$ experimentally. Figure~\ref{fig:fig2}(b) shows the domain-wall velocity as a function of the net spin density $\delta_s$ within the phase-locking regime for the frequency $f = \omega/2\pi = 90$ MHz. The analytical solutions [Eq.~(\ref{eq:V1})] and the simulation results are depicted by square and triangle symbols, respectively. Note that the sign of the velocity changes, i.e., the direction of the domain-wall motion reverses, as the net spin density changes its sign. The results with all the temperatures except the magnetization compensation point $T_4(T_M)$ are used~\footnote{The frequency $f = 90$ MHz belongs to the phase-unlocking regime for the magnetization compensation point, and thus it is excluded from Fig.~\ref{fig:fig2}(b).}.

\begin{figure*}
\includegraphics[width=\textwidth]{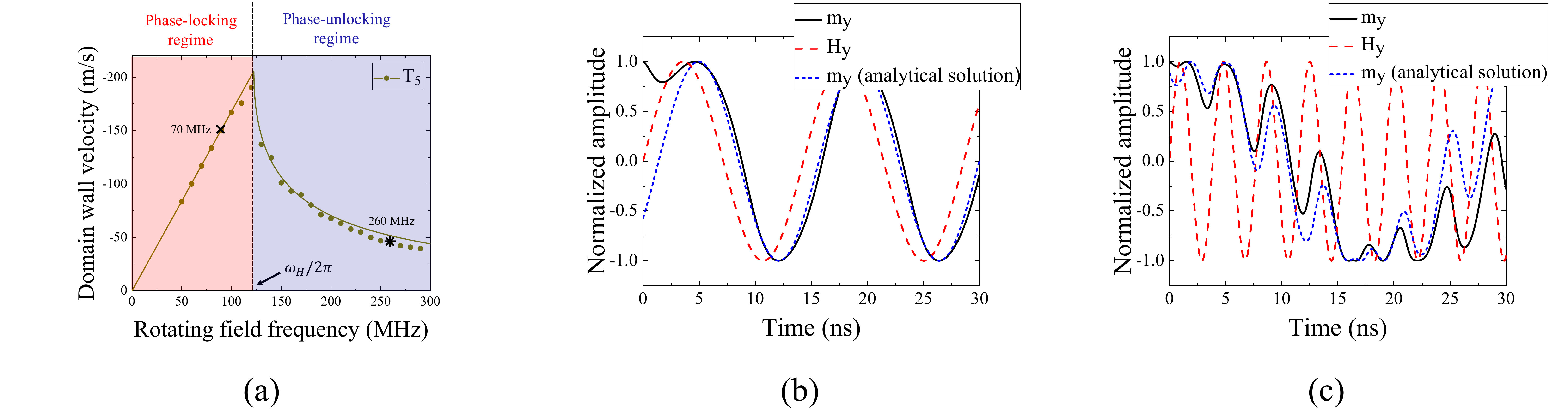}
\caption{(a) Domain-wall velocity $\langle \dot{X} \rangle$ as a function of the rotating-field frequency $f = \omega / 2 \pi$ for the case of $T_5$ (defined in Table~\ref{table}). The lines and the dots represent the analytical solutions and the simulation results, respectively. The critical frequency $\omega_H$ which separates the two distinct regimes of domain-wall dynamics is calculated from Eq.~(\ref{eq:wh}) and is shown as the vertical dashed line. When the frequency is below the critical frequency, the dynamics of the domain wall is in the phase-locking regime, where the domain wall precesses at the same frequency of the rotating field and thus the velocity increases linearly as a function of the frequency. When the frequency is above the critical frequency, the dynamics of the domain wall motion is in the phase-unlocking regime and the resultant velocity decreases as the frequency increases. (b, c) Time evolution of the $y$-component of the magnetization $m_y$ at the domain-wall center and the external field $H_y$ for the rotating-field frequency of (b) 70 MHz and (c) 260 MHz. The black solid line and the red dashed line are obtained from the simulations. The blue dotted lines are the analytical solutions obtained from (b) Eq.~(\ref{eq:sol1}) and (c) Eq.~(\ref{eq:sol2}).}
\label{fig:fig3}
\end{figure*}

Figure~\ref{fig:fig3}(a) shows the domain-wall velocity as a function of the frequency for the configuration $T_5$ (see Table~\ref{table} for the definition). The critical frequency $\omega_H$ obtained from Eq.~(\ref{eq:wh}) is shown as a vertical dashed line. Figure~\ref{fig:fig3}(b) and (c) show the evolution of $m_y(t)$, the $y$-component of the magnetization evaluated at the domain-wall center and $H_y(t)$, the $y$-component of the rotating field at the frequencies of $70$ MHz and $260$ MHz, respectively. The black solid lines and the red dashed lines represent the simulation results for $m_y$ and $H_y$, respectively. The dashed blue lines show the analytical solutions given by (b) Eq.~(\ref{eq:sol1}) and (c) Eq.~(\ref{eq:sol2}). In the phase-locking regime shown in Fig.~\ref{fig:fig3}(b), the domain-wall magnetization precesses at the same frequency of the external field and thereby the domain-wall velocity increases linearly as the frequency increases as expected from the analytical solution [Eq.~(\ref{eq:V1})]. In the phase-unlocking regime shown in Fig.~\ref{fig:fig3}(c), the time duration for $m_y$ to change between $1$ and $-1$ is much longer than the period of $H_y$, meaning that the domain-wall precession is much slower than the applied field. In this case, the domain-wall velocity decreases as the frequency increases [Eq.~(\ref{eq:V2})]. There are some deviations between the analytical solution and the simulation result in Fig.~\ref{fig:fig3}(a, b), which are presumably due to the approximations that we take to obtain the analytical results such as neglecting $\ddot{X}$ and $\ddot{\Phi}$ in Eq.~(\ref{eqn3}) and Eq.~(\ref{eqn4}).

\section{Summary}
\label{sec4}

We have studied the dynamics of a ferrimagnetic domain wall driven by a rotating field analytically by using the Landau-Lifshitz-Gilbert-like equations for ferrimagnets and also by numerically solving the coupled LLG equations. We have found that, depending on the frequency of the field rotation, there are two distinct regimes of the dynamics of a domain wall. In the phase-locking regime, where the frequency is below the critical frequency, the domain-wall velocity is proportional to the frequency of the rotating field. In the phase-unlocking regime where the frequency is above the critical frequency, the domain-wall velocity decreases as the frequency increases. In addition, we have found that the direction of the domain-wall motion depends on the sign of the net spin density of the ferrimagnet. This results in the reversal of the domain-wall velocity sign as the temperature varies across the angular momentum compensation point $T_A$.

\begin{acknowledgments}
This work was supported by Brain Pool Plus Program through the National Research Foundation of Korea funded by the Ministry of Science and ICT (Grant No. NRF-2020H1D3A2A03099291), by the National Research Foundation of Korea funded by the Korea Government via the SRC Center for Quantum Coherence in Condensed Matter (Grant No. NRF-2016R1A5A1008184). K.J.L. was supported by the National Research Foundation of Korea (Grant No. NRF-2015M3D1A1070465). D.H.K was supported by the POSCO Science Fellowship of POSCO TJ Park Foundation, by the Korea Institute of Science and Technology (KIST) institutional program (No. 2E31032), and by the National Research Council of Science \& Technology (NST) grant (Project No. 2N45290) funded by the Korea government (Ministry of Science and ICT).
\end{acknowledgments}

\bibliography{ref.bib}

\end{document}